\documentclass[conference,compsoc]{IEEEtran}
\ifCLASSOPTIONcompsoc
\usepackage[nocompress]{cite}
\usepackage[cmintegrals]{newtxmath}

\else

  \usepackage{cite}
\fi

\ifCLASSINFOpdf
  \usepackage[pdftex]{graphicx}

  \graphicspath{{../pdf/}{../jpeg/}}

\else

\usepackage[dvips]{graphicx}
  \graphicspath{{../eps/}}
  
\fi
\usepackage{amsmath}
\makeatletter
\let\@tmp\@xfloat     
\usepackage{fixltx2e}
\let\@xfloat\@tmp 
\usepackage{array}

\ifCLASSOPTIONcompsoc
 \usepackage[caption=false,font=footnotesize,labelfont=sf,textfont=sf]{subfig}
\else
  \usepackage[caption=false,font=foot\fi
\usepackage{fixltx2e}

\usepackage{stfloats}
\usepackage{url}

\begin{document}

\title{An adaptive self-organizing fuzzy logic controller in a serious game for motor impairment rehabilitation}

\author{\IEEEauthorblockN{Shabnam~Sadeghi~Esfahlani,~Silvia~Cirstea, George~Wilson and Alireza~Sanaei}
\IEEEauthorblockA{Department of Computing and Technology\\ Anglia Ruskin University, Cambridge, UK\\
\\
Email: shabnam.sadeghi-esfahlani@anglia.ac.uk, \\silvia.cirstea@anglia.ac.uk,
george.wilson@anglia.ac.uk, alireza.sanaei@anglia.ac.uk}}

\maketitle

\begin{abstract}
Rehabiliation robotics combined with video game technology provides a means of assisting in the rehabilitation of patients with neuromuscular disorders by performing various facilitation movements. The current work presents ReHabGame, a serious game using a fusion of implemented technologies that can be easily used by patients and therapists to assess and enhance sensorimotor performance and also increase the activities in the daily lives of patients. The game allows a player to control avatar movements through a Kinect Xbox, Myo armband and rudder foot pedal, and involves a series of reach-grasp-collect tasks whose difficulty levels are learnt by a fuzzy interface. The orientation, angular velocity, head and spine tilts and other data generated by the player are monitored and saved, whilst the task completion is calculated by solving an inverse kinematics algorithm which orientates the upper limb joints of the avatar. The different values in upper body quantities of movement provide fuzzy input from which crisp output is determined and used to generate an appropriate subsequent rehabilitation game level. The system can thus provide personalised, autonomously-learnt rehabilitation programmes for patients with neuromuscular disorders with superior predictions to guide the development of improved clinical protocols compared to traditional theraputic activities.
\end{abstract}
\IEEEpeerreviewmaketitle
\section{Introduction}
Computational intelligence (CI) and artificial intelligence (AI) can offer unsupervised and hybrid instruction that enables relearning sensorimotor function without the need of direct supervision \cite{lai2009computational} and \cite{xie2011iterative}. Video games and virtual reality (VR) systems when combined with AI and CI algorithms provide the possibility of designing and developing an environment through which the well-being and mobility of a person could be maintained. More generally, video games can be employed for physiotherapy as well as pain management to distract awareness \cite{kavalier2005video}. Unlike more traditional therapeutic activities, therapy using VR or video games does not rely on passive movements that could sometimes accompany painful manipulation of the limbs.\\
It is well known that a person can manage a complicated process on his/her own using heuristics whereby the expert knowledge is acquired by experimentation or trial and error without the aid of any closed control loop  \cite{lilly2011fuzzy}. Actions based on the expert knowledge requires sophisticated methods to justify the non-linearity and complexity of real-world problems, especially when these issues could not be addressed through sufficient theories, mathematical or analytic models \cite{czekalski2006evolution}. 
Systems with complex non-linear control problems require highly sophisticated computer systems on which to build the knowledge \cite{lilly2011fuzzy}. In general, it is of vital importance to design a framework that addresses more efficient and effective diagnostic and rehabilitation methods \cite{lai2009computational}.
This structure can be implemented using a fuzzy logic that utilises a linguistic system by employing the terms of natural language and makes decisions based on prior knowledge of the scheme \cite{zarandi2008reinforcement}. It is one of the great strengths of fuzzy control that expert knowledge is incorporated in the absence of mathematical theories. This is capable of reflecting reality and able to generalise the experience in uncertain cases. The Mamdani-Assilian linguistic influence system is arguably the best known method, which is based on the fundamental work of Zadeh (1971) \cite{czekalski2006evolution}, \cite{zadeh1994role} and \cite{gaines1972learning}. \\
In this study, an integrated system called 'ReHabGame' is developed for body posture monitoring and physical therapy rehabilitation. The framework is based on a Kinect Xbox One and Myo armband skeleton and joint tracking devices and a rudder foot pedal that enables navigating the virtual world. The adaptive rehabilitation system is built on robotic rehabilitation techniques, inverse kinematics (IK), and Mamdani's fuzzy logic method. The decisions are made by the system based on the player's performance and his/her level of physical ability. The data is extracted and analysed via Mamdani's fuzzy logic according to the sequence of postures and orientations of the player and the virtual robotic instructor in the 3D environment.
ReHabGame is a serious game designed specifically to cater for the issues of a particular therapeutic task and involves the development of four different game scenarios. Real-time feedback and other data are provided through the Kinect Xbox, Myo armband, and Rudder Pedal devices. The system is developed through the device interfaces and the 3D virtual games via the Unity3D game engine. The recorded and measured quantities are imported as input into a fuzzy rule system for fuzzification. Every rule has a weight (between 0 and 1) from which the implication method is implemented.
Defuzzification is then performed to convert the collection of data into a crisp output. These output data are presented as "progress into the next level, repeat the level, make the level more simple or the activity was harmful" and are fed back into Unity. If the crisp output suggests progression to a next level, the game shifts to a more advanced stage. The work presented here shows the effectiveness of machine learning techniques in refining physiotherapy arrangements in a way to achieve a desirable outcome. 
\subsection{Background \& Related Work}
The development and simulation of fuzzy logic based learning mechanisms related to robotic rehabilitation is well-documented, utilizing various devices to emulate human motor learning \cite{xie2011iterative}, \cite{macdonald2007formation}, \cite{singh2013real}, \cite{nawrocka2014fuzzy}, \cite{skoda2015estimation}, \cite{cho2015control}, and \cite{huq2013development}.\\
The relationships between movements of upper extremities identifying joint angles using a fuzzy logic system have been reported \cite{chang2008experimental}; these authors analysed a range of joint angles and rhythmical movement variables to design a fuzzy expert system. \cite{macdonald2007formation} utilised fuzzy inference to develop an internal model of a dynamic environment experienced during planar reaching movements with the upper limb. They considered trajectory movement and velocity errors as input variables to update parameters and showed that the fuzzy learning strategy is successful in incrementally updating an internal model of the environment dynamics. 
\cite{karime2014fuzzy} have introduced and assessed an adaptive home based rehabilitation framework for wrist training. A fuzzy logic adaptation engine was used and evaluated while coping with different levels of exercise. A physical therapy rehabilitation serious game was developed by \cite{fernandez2015serious} and is a 2D game based on the Kinect tracking skeleton in which the authors used a fuzzy logic method to analyse the sequence of player's postures which could be compared against a specification of the prescribed exercise. A self-adaptive home based rehabilitation game has also been developed by \cite{pirovano2012self}, who  present a 2D game where gameplay is continuously monitored using a fuzzy system. They showed that computational intelligence could offer a feasible and robust method of home rehabilitation.  
\cite{liu2014fuzzy} have proposed a fuzzy logic risk-predictor of Parkinson's disease through a 2D game called Pumpkin Garden. The game takes a player's historical in-game finger movement and other behaviour data as input and calculates the player's risk of developing the disease accordingly. \\
\section{Skeleton joints and their hierarchy}
\begin{figure}[!t]
\centering
\includegraphics[width=3.6 in]{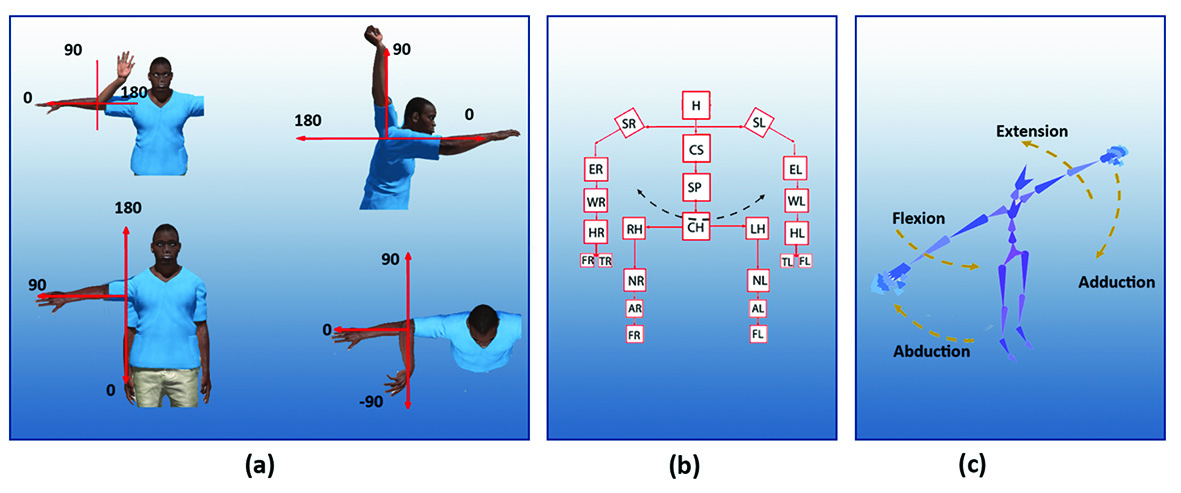}
\caption{Skeleton and body orientation, joints and their hierarchy.}
\label{Fig1}
\end{figure}

The Kinect skeleton tracking system detects the position and orientation of individual joints and the speed of joint movement.
The upper limb orientation with the maximum and minimum values are shown in Fig. \ref{Fig1}-a. The joint hierarchy is illustrated in Fig. \ref{Fig1}-b. The joints have their origin in the Spine (SP) and extend downwards to the centre of the hip (CH) and from there to the right hip (RH) and the left hip (LH). The SP extends upwards to the middle of the shoulders (CS) and head (H), towards the right and left shoulders (SR) and (SL), elbow (ER) and (EL), wrist (WR) and (WL), and hand (RH) and (LH).
Fig. \ref{Fig1}-c shows the terms used for upper limb movements. The upper arm abduction (moves the limb laterally away from the midline of the body), adduction (the opposing action that brings the arm toward the body or across the midline), flexion (decreases the angle between the bones or bending of the joint), and extension (increases the angle and straightens the joint).
\section{Kinect, Myo Armband, and Saitek Pro Flight Rudder Pedal}
\begin{figure}[!t]
\centering
\includegraphics[width=3.5in]{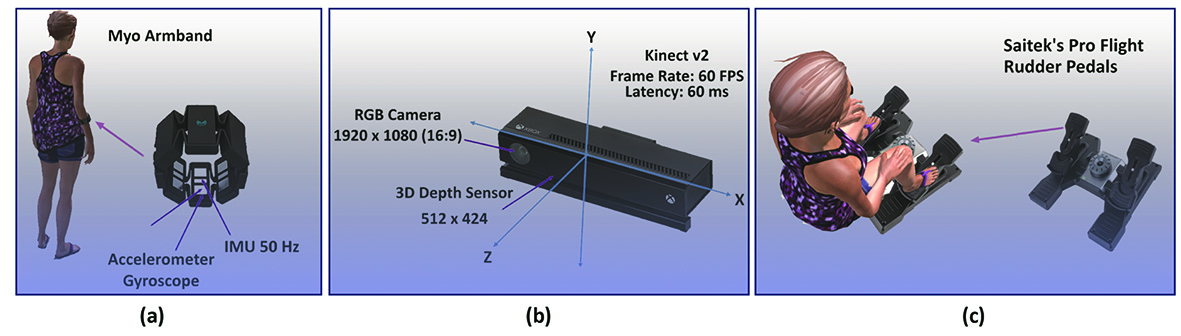}
\caption{Devices that are interfaced with the rehabilitation game via Unity3D.}
\label{Fig2}
\end{figure}

Microsoft Kinect Xbox One version 2 (Microsoft in 2010) and Myo armband (Thalmic Labs in 2016) devices offer a portable 3D motion capture capability, enabling users to control and interact with a computer in real time. Kinect V2 is illustrated in Fig.~\ref{Fig2}-b and consists of an infrared laser based IR emitter and a colour (RGB) camera.  It has the capability to detect the position and orientation of $25$ individual joints, the speed of player's joint and body movements, as well as to  track gestures performed with a standard controller.\\
The upper limb is modeled as a kinematic chain consisting of: three degrees of freedom (DoF) for the shoulder spherical joint (i.e. abduction-adduction, flexion-extension and internal-external rotation of the upper arm), one DoF for the revolute elbow joint (i.e. forearm flexion-extension, indicated), three DoF for the wrist spherical common (i.e. pronation-supination, ulnar-radial deviation and flexion-extension of the hand). 
The Myo armband is shown in Fig.~\ref{Fig2}-a; it streams the accelerometer, gyroscope, and electromyography (EMG) data and the inertial measurement unit (IMU) data at $200$ $Hz$ and $50$ $Hz$ frequency, respectively. The three-axis accelerometer, magnetometer and gyroscope senses and records motion and collects real-time data with a high accuracy and precision. It is made of eight medical grade stainless steel EMG sensors that detect the electric impulses in the muscles. The armband is connected via Bluetooth USB adapter and interacts with the VR through the movements of the hand and wrist. Underlying electrical activity in the muscle is detected (electromyography) and displayed as a waveform (an electromyogram).\\
The Saitek Pro Flight Rudder Pedal\footnote{http://www.saitek.com/uk/prod/pedals.html} with 3-axes is used for navigation inside the virtual 3D environment and to simulate movements. The movement is controlled by the player while he/she is sited and pushing the pedals in alternative sequences for forward movement, and left or right toe push for rotating towards left or right. It is connected to the game engine and virtual game via a USB connector and has an adjustable tension dial that is used to select a suitable resistance in order to allow exertion of an appropriate pressure within a range of forces. The differential braking provides independent control of the left and right toe brakes that are used for moving left and right. 
The footrests adjust to accommodate a range of sizes and include non-slip materials to help keep the pedals on the floor (Fig.~\ref{Fig2}-c).\\
\section{Kinematics of leading joints and Inverse Kinematics}
\begin{figure}[!t]
\centering
\includegraphics[width=3.2in]{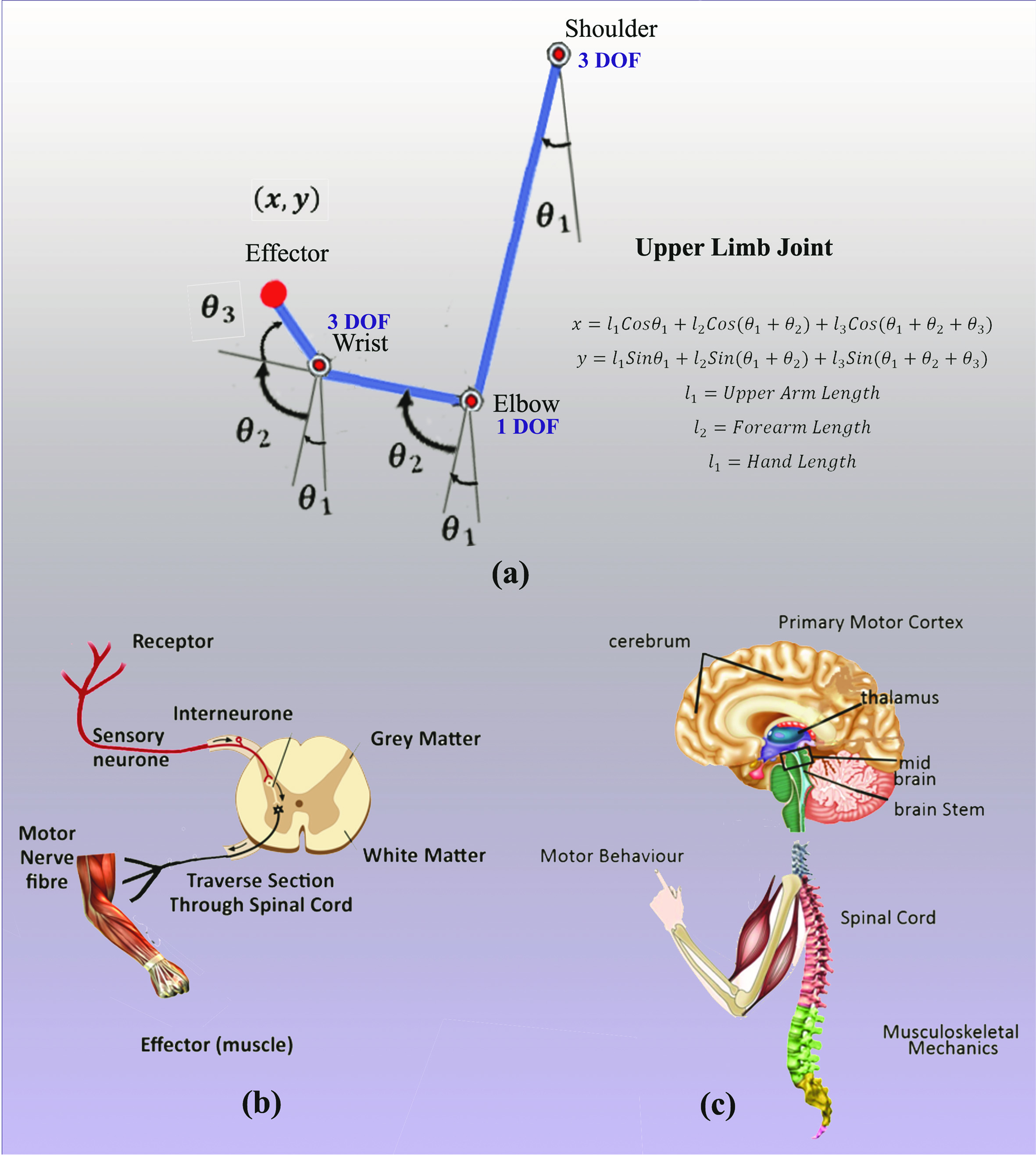}
\caption{Spinal reflex arc (a) and organisation in the brain and its sensory neurones (b)}
\label{nerves}
\end{figure}

Kinematics refer to the mathematical description of motion without considering the underlying physical forces. The kinematics of the human body is specifically concerned with formulating and solving for the translation, rotation, position and velocity of
each body segment in real-world motions \cite{grecu2009analysis}.
The movement of a kinematic chain, whether it is a robot or an animated character, is modeled by the kinematic equations of the chain \cite{grecu2009analysis}, \cite {papaleo2012inverse}, \cite{tolani1996real} and \cite{lura2012creation}.
The IK determines the typical parameters that provide a desired position of the end-effector to achieve the task and is known as action planning \cite{lura2012creation}.
IK techniques provide direct control over the placement of an end effector object at the end of a kinematic chain of joints, solving for the joint rotations which place the object at the desired location. IK offers an alternative to explicitly rotating individual joints within a skeleton \cite{papaleo2012inverse}, \cite{tolani1996real}, and \cite{welman1993inverse}. As such, the virtual robotic avatar is animated using IK in that the position of an end effector is presented to the algorithm so that the system automatically computes the joint angles needed to collect the object (Fig~\ref{nerves}-a). The IK equation is defined in Eqn. \ref{InverseKin}.\\
The data collected from Kinect, Myo armband and Rudder foot pedal devices are used to monitor a player's performance. Below is the list of parameters and data that can be collected from these devices:\\
 \begin{equation}
\label{InverseKin}
\begin{array}{llllll}
\theta_1\leftarrow \arctan2(y,x)\\
\theta_2\leftarrow \arctan2(z-l_1 , \sqrt[]{x^2+y^2})-\arctan2(l_3s_3 , l_2+l_3c_3) \\
\theta_3\leftarrow \arctan2(s_3,c_3)\\
\\
where:  \left\{
\begin{array}{ll}
c_3\leftarrow \frac{x^2+y^2+(z-l_1)^2-l_2^2-l_3^2}{2l_2l_3}\\\\
s_3\leftarrow  \sqrt[]{1-c_3^2}
\end{array}
\right.
\end{array} 
 \end{equation}

\textbf{Data taken from Kinect XBox One}
 \begin{itemize}
\item The orientation (O) and position (P) of the wrist ($O_{W}$), ($P_{W}$) the elbow ($O_{E}$), ($P_{E}$), and the shoulder ($O_{S}$), ($P_{S}$) when  the objects are spawned relative to their final location (effector).\\
\item Average angular velocity (AV) of the  wrist($AV_{W}$), the elbow ($AV_{E}$), and the shoulder ($AV_{S}$). It is defined by the Eqn.~\ref{Eqn. 2}, where $v(t)$ is the angular velocity, $t_1$ is the starting time of the task and $t_2$ is the time the task is finished~\cite{karime2014fuzzy}.
 \begin{equation}
\label{Eqn. 2}
AV=\frac{1}{t_2-t_1}\int^{t_2}_{t_1}v(t)dt
\end{equation}
\item The time an object is "spawned", "reached" and "collected".\\
\item Head Tilt $(T_H)$ and Spine  Tilt $(T_S)$ are used to measure the postural body stability.
\end{itemize}
\textbf{Data taken from Rudder Foot Pedal}
\begin{itemize}
\item Pressure on Foot Pedal: The amount of pressure applied by each foot are monitored. It ranges between $[-1,1]$ to determine (I) if the steps are taken in order (otherwise the forward movement would not take place), (II) measure the amount of pressure applied on each pedal;
Left foot pressure $P_{(LF)}$ is $-1$, neutral $P_{(N)}$ is $0$ and right foot pressure $P_{(RF)}$ is $+1$. The data is recorded per frame, and the average value is calculated $P_{(Avg)}$. If the $P_{(Avg)}$ is negative then this shows the left foot is dominant and if it is positive the right foot is the dominant one. 
\end{itemize}
\textbf{Data taken from Myo Armband} 
\begin{itemize}
 \item Eight ElectroMyoGraphy (EMG) data sets are collected to evaluate the electrical activity produced by the muscles of the upper or lower arm. These data are saved individually as $EMG_1$, $EMG_2$, $EMG_3$, $EMG_4$, $EMG_5$, $EMG_6$, $EMG_7$, and $EMG_8$.\\
\item The gyroscope measures the rotation around one of the axes (X, Y, and Z). When it is combined with the accelerometer, it provides a  useful, compact measurement system, known as inertial measurement unit (IMU). The data is used to measure the arm movements and its activities when playing the game (motion, speed, times, vertical and horizontal displacements as well as kinematic parameters). Data is recorded as $IMU_X$, $IMU_Y$, and $IMU_Z$.\\
 \end{itemize}
\section{Adaptive Rehabilitation Games}
In designing the ReHabGame, a broad spectrum of rehabilitation exercises were studied with advice obtained through direct collaboration and consultancy sessions with physiotherapy experts. As such, the involvement of various body segments is considered in developing the games to reach the desired goals. Fig.~\ref{nerves} (b) and (c) show the spinal reflex arc, brain stem and the musculoskeletal mechanics of the body. As shown, the brain has the important role in transferring the signals from muscle and motor nerve fibre to sensory neurones and receptors. Accordingly, data is transferred back to the spinal cord to trigger motor behaviour or any action. Any impairment in this system could effect the data transfer between various segments and as a result could cause reduced functionality of the sensitive parts and connectors. Thus, the adaptive ReHabGame is developed to assist with various sensorimotor functions considering the following factors: 
\begin{itemize}
\item spinal reflexes (reflex of the joint during abnormal stress condition) and brain stem activity (cognitive awareness of voluntary movements).
\item Postural stability and balance in standing and walking. It involves the control of position and motion of the centre of body mass about the base of support. In upright posture, instability may cause the whole body to slightly sway while standing still, so there is a constant subtle fluctuation between stability and mobility to maintain balance.
\item Residual cognitive impairment and deficiencies in the upper and lower limb movements.
\end{itemize}
The rehabilitation therapy is performed through four game scenarios \cite{esfahlani2016intelligent}. Fig~\ref{Menu}-I shows the main menu of the game in that the patient or player's data must be entered and will be saved in a folder before choosing any of the games. This folder accumulates the player's activity during the game session with the exact times, dates, joints and muscles data. Fig~\ref{Menu}-II illustrates the second menu with different quantities that are used for parametrizing the constraints. The parameterized quantities are; maximum and minimum range of shoulder, elbow and hand orientation and position, the right-handed or the left-handed game, the object generator algorithms, and some iterations, the basket size, fruit size, and the speed of fruit that are spawned. The four scenarios of the ReHabGame are listed below:
\begin{itemize}
\item \textbf{"Reach-Grasp-Collect fruits"} game to engage the upper limbs and body posture that is a "Two Players" game as depicted in Fig~\ref{Menu}-a and the "Single Player" game is shown in Fig~\ref{Menu}-b. The player/s should reach the fruits generated in the scene to grab and collect them in the designated basket and gain score. 
\begin{figure*}[!t]
\centering
\includegraphics[width=5.0in]{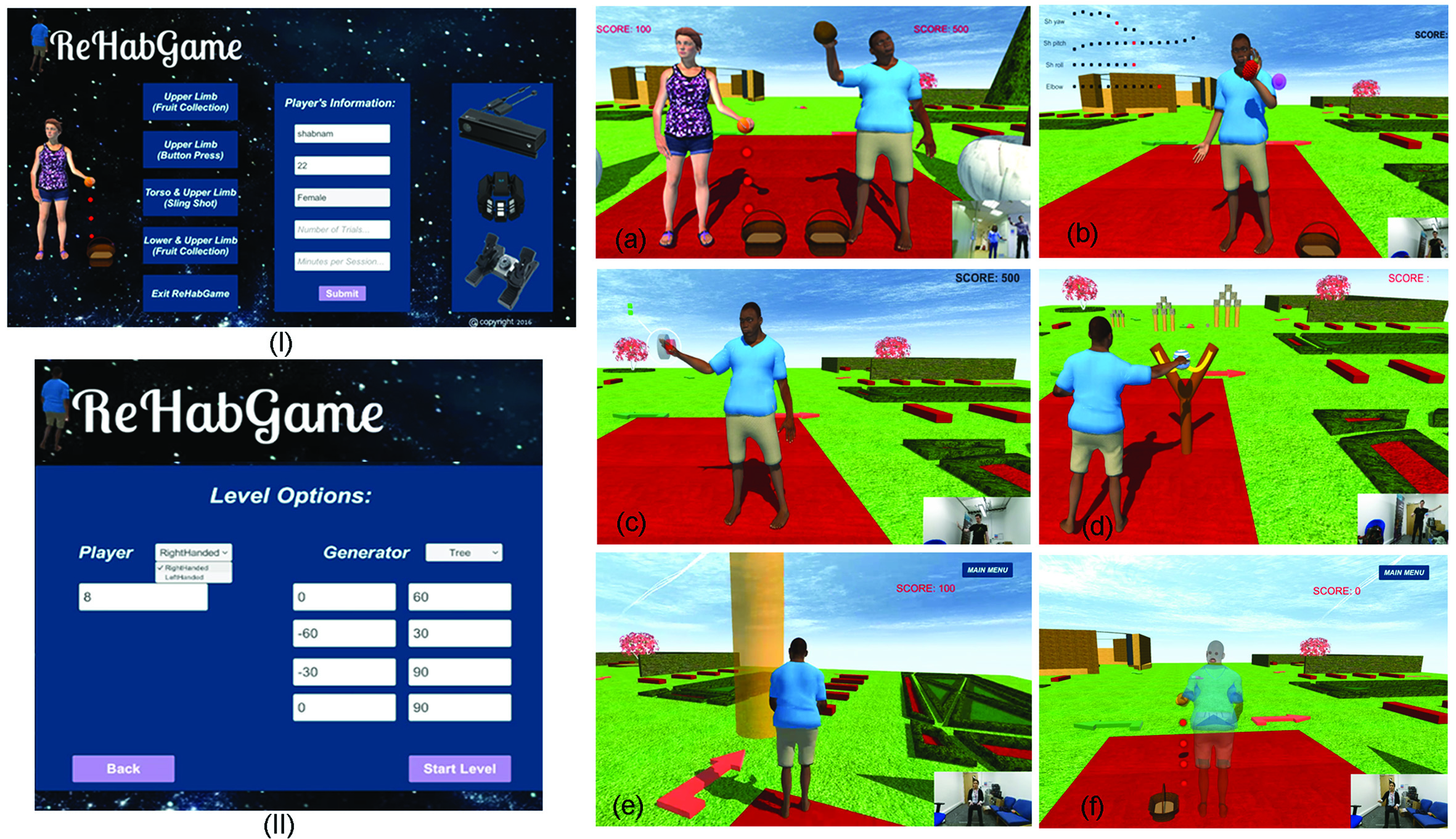}
\caption{The main menu of the ReHabGame (I) second menu to adjust constraints (II), "Two players" game (a), "Single player" game (b), Button press and hold game (c), Slingshot/hit game (d) and maneuvering in the scene and fruit collection in the virtual world game (e) and (f).}
\label{Menu}
\end{figure*}
\item \textbf{"Button Reach-Press-Hold"} game in that buttons are generated in a matrix format. The player should reach the buttons and hold them steadily for a particular amount of time or just reach and press it Fig~\ref{Menu}-c.
\item \textbf{"Sling Shot and Hit"} game for engaging torso and upper limb. The player should pull the sling stretching it with the aim of hitting the target cubes on release Fig~\ref{Menu}-d.
\item \textbf{"Manoeuvring and Reach-Grasp-Collect fruits"} game to stimulate lower and upper limb activity. It is done by manoeuvring in the 3D VR scene. The game is interfaced with the foot pedal to simulate the walking of the avatar. The player must reach the highlighted random areas in the VR using the foot pedal, attain the fruits generated and collect them in the basket. The player must reach those areas and collect the rewards. 
\end{itemize}
\begin{figure}[!t]
\centering
\includegraphics[width=3.5 in]{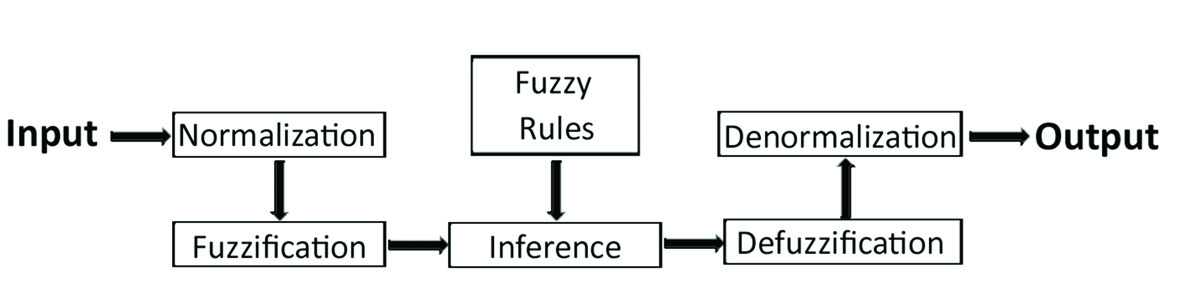}
\caption{General schematic of fuzzy regulation.}
\label{Fuzzy}
\end{figure}
\subsection{Fuzzy Logic and Machine Learning Algorithm}
The fuzzy interface is adapted first to "Reach-Grasp-Collect fruit" game and then is expanded to other games. The fuzzy logic adaptation system makes decisions based on the player's performance and accordingly proposes an appropriate level of rehabilitation game. It is done based on human's logic that mimics the same reasoning strategy
 \cite{lilly2011fuzzy} and \cite{wang1997course}. The intensity of the game and game progression is based on completion of a task rather than by per-session. In the fuzzy logic, there is a gradual transition from one value to another rather than abrupt change (granulation). There are overlapping fuzzy interval changes that mimic the way in which human perceive linguistic reasoning. Fuzzy logic provides a framework for the description of imprecise dependencies and commands through the use of fuzzy $if/then$ or $or/and$ rules. The concepts which play central roles in the application of fuzzy set are linguistic variable and values, membership functions, and the universe of discourse \cite{zadeh1994role}. A fuzzy set is a collection of real numbers having partial membership in the set. The total number of members in the set is specified by a membership value of $1$, absolute exclusion from the game has a member value of $0$ and partial membership in the set is determined by a membership value between $0$ and $1$ \cite{lilly2011fuzzy}. A membership function (MF) is a curve that defines how each point in the input space is mapped to a membership value in $[0,1]$. The role of the membership functions is to find the standard range for each input  \cite{lilly2011fuzzy}, \cite{macdonald2007formation},
\cite{chang2008experimental}, \cite{chang2010adaptive} and \cite{barghout2009haptic}, MF specifies a fuzzy set $A$ as Eqn. \ref{Eqn. 3},
\cite{czogala2012fuzzy}.

\begin{equation}
\label{Eqn. 3}
\mu : X \rightarrow [0,1] ~or~ A:X \rightarrow [0,1].
\end{equation}
A fuzzy set $A$ in $X$ is directly specified by the function $(x,\mu_A(x))$, mathematically as Eqn. \ref{Eqn. 4}:  
\begin{equation}
\label{Eqn. 4}
A=\{(x,\mu_A(x)) \mid x \in X\}
\end{equation}
The collection of numbers on which a variable is defined are the universe of discourse for the variable. Consider the system where a variable within a universe of discourse is $X$, and $x$ is a real number where ($x \in X$) and let $A$ denote a fuzzy set defined on $X$. For these conditions a membership function $\mu_A (x)$ associated with $A$ is a function that maps $X$ into $[0,1]$ and gives the grade of membership of $X$ as $A$. The MF that is used for the ReHabgame fuzzy logic system is triangular with the only condition that it must be in the range of $[0,1]$ \cite{lilly2011fuzzy}. The triangular membership function is defined by Eqn. \ref{Eqn. 5} where $\alpha$, $\beta$, and $\gamma$ are the parameters. The fuzzy logic operates on the input $x$ to produce the crisp output $\mu$.
\begin{equation}
\label{Eqn. 5}
\mu_A(x:\alpha,\beta,\gamma)=
 \left\{
 \begin{array}{llll}
0 \  \  \ \ \ \  \  \ \ x \leqslant \alpha\\
\frac{x-\alpha}{\beta-\alpha}  \ \ \ \alpha < x \leqslant \beta\\
\frac{\gamma - x}{\gamma - \beta}  \ \ \ \beta \leqslant x < \gamma\\
0 \  \  \ \ \ \  \  \ \ x \geqslant \gamma \\
              \end{array}
              \right.
\end{equation}
In this study a multi-input and single-output (MISO) fuzzy system is developed \cite{lilly2011fuzzy}. The MISO model is designed with $n$ inputs and every $i^{th}$ fuzzy if-then rule ($R^{i}$) is specified by a set of fuzzy input sets $A^{i}_n$, an output fuzzy set $B^{i}$ and a set of parameters $\theta$ that can be represented using Eqn. \ref{Eqn. 6} \cite{czogala2012fuzzy} and \cite{lilly2011fuzzy}:
\begin{equation}
\label{Eqn. 6}
R^{i}: \textbf{\textit{IF}}~ 
{and_{n=1}^{N}}
~X_n~ \textbf{\textit{is}}~A^{i}_n ~\textbf{\textit{THEN}}~\mu~\textbf{\textit{is}}~B^{i}(\theta,x_0)
\end{equation}
$X_n$ and $\mu$ denotes linguistic variables (inputs and output, respectively), and $B^{i}(\theta,x_0)$ represents a parameterized linguistic value of the output linguistic variable $\mu$. \\
Each triangular function is composed by a set of $if-then$ rules as in Eqn. \ref{Eqn. 7}.
\begin{equation}
\label{Eqn. 7}
 \left\{
\begin{array}{lll}
$If $ ((x \leqslant \alpha) \land (x\geqslant\gamma)) \Rightarrow \ \ \ f(x)= 0 \\
$If $ ((x > \alpha) \land (x \leqslant \beta)) \Rightarrow \ \ \ f(x)= \frac{x -\alpha}{\alpha - \beta} \\
$If $ ((x \geqslant \beta) \land (x < \gamma)) \Rightarrow \ \ \ f(x)= \frac{\gamma - x}{\gamma - \beta} \\
 \end{array}
  \right.
\end{equation}
According to the schematic of the fuzzy system illustrated in Fig. \ref{Fuzzy} in order to get crisp output from the system the defuzzification interface is required in that the contribution of each fuzzy set inferred is individually considered by means of a characteristic value center of gravity and the final crisp value is obtained by means of a weighted average aggregation operator \cite{cordon1997applicability} and \cite{alcala1999techniques}.

\subsection{Fuzzy Logic Controller in the ReHabgame}
Two main tasks are needed to design an intelligent system for the ReHabgame. Firstly the fuzzy operators involved in the Inference System need to be selected, and secondly adequate knowledge about the problem being solved needs to be derived \cite{alcala1999techniques}, \cite{bigler2012mild}, \cite{dounskaia2010control}, and \cite{siler2005fuzzy}.
The intelligent fuzzy logic inference system proposes an exercise that is more appropriate for the player based on his/her performance with out any further harm. The normality factor of the training is adopted on input parameters that provides a quantified value that describes how well a patient performed in a particular task \cite{fernandez2015serious} and \cite{liu2014fuzzy}. Input quantities are standardised and transferred for fuzzification through fuzzy sets, a set of predefined rules are applied to fuzzy sets through the fuzzy inference that moves them through defuzzification and denormalization step to the crisp output $\mu$.\\ Below is the list of steps towards designing the fuzzy logic system:\\
\textbf{Input and output variables:} The input variables that are used depend on the subjective assessment and the severity of the constraints in the games. A maximum of thirteen fuzzy variables (inputs) are considered depending on each specific task and the constraints and potentially generate an output which directs the player towards Progression", Repetition", "Simplification" or "Harmfulness". This input data can be activated or deactivated upon request by the clinician or the player before attempting the game depending on the constraints that need to be applied. Four subsets are assigned to the input variables position, orientation and average angular velocity as follows: "Very Good (VG), Good(G), Bad (B), and Harmful (H)". The timing subsets are "Very Good (VG), Good(G), Bad (B)". Some fuzzy logic input functions are described below along with the output function.
\begin{itemize}
\item \textbf{$O.E_W$, ($P.E_W$), $O.E_E$, ($P.E_E$) and $O.E_S$ ($P.E_S$): The Orientation (and Position) deviation (error) of the Wrist, Elbow and Shoulder}, defined as the difference between the player controlled avatar joints orientation (and position)  against virtual robotic avatar. The orientation subsets errors change within the range of $[0^{\circ},90^{\circ}]$ where $0^{\circ}$ is "VG" and $90^{\circ}$ is "H". and the positions are within $[0,90]$. Where $0$ is "VG" and $90$ is "H". \\
\item \textbf{$AV.E_W$, $AV.E_E$, and $AV.E_S$: The Average Angular Velocity Error of the Wrist, Elbow and Shoulder}. Defined as the difference between the joints average angular velocities. The element of a universe of discourse is in the range of $[0,90]$ where $0$ is "VG" and $90$ is "H".\\
\item \textbf{$T.E_H$ and $T.E_{S}$: The Head and spine Tilt Error}. These parameters monitor the player's upright pose (the body posture) in that the spine tilt error $T.E_S$ is in the range of $[0^{\circ}, 36^{\circ}] \deg$ in that $36^{\circ}$ is "H". The head inclination is in the range of $[0 ^{\circ},32^{\circ}]$ in each direction where $0^{\circ}$ is "VG" and $32 ^{\circ}$ is "H".\\
\item \textbf{$T.E_C$ and $T.E_R$: The Collection and Release Time Error}, defined as the difference between the collection and release time. These two variables are in the range of $[0,6]$ where $0$ is "VG" and $6$ is "B".\\
\item \textbf{The output variable "GameProgress":} with four subsets ranging in $[0,80]$. Where $[0-20]$ is "Progression", $[20-40]$ is "Repetition", $[40-60]$ is "Simplification" and $[60-80]$ is "Harmfulness".
\begin{figure}[!t]
\centering\includegraphics[width=3.2in]{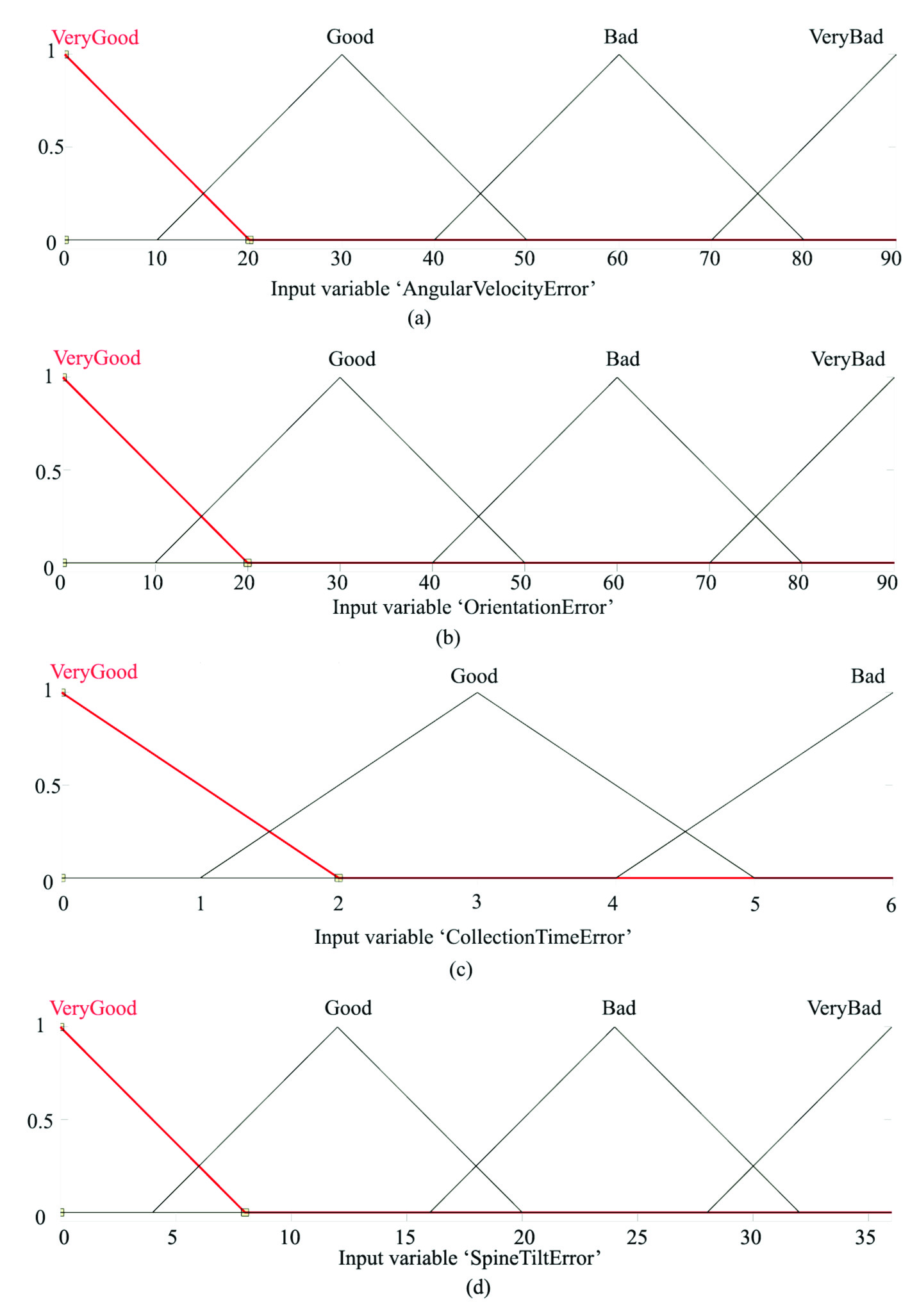}
\caption{Some examples of fuzzy logic input variables defined in the ReHabGame.}
\label{inputs}
\end{figure}
\end{itemize}
The triangular fuzzy set diagrams are consistent within the constraints of the above conditions with small arbitrary error overlaps (Fig. \ref{inputs}). This follows a standard practice for fuzzy set design of a new system and simplifies the calculations so reducing the demand on computer resources.\\ 
\textbf{Linguistic Rules and Constrains:} 
The control rules are implemented using fuzzy conditional statements (if/then, or/and, and not) and the relations between subsets. $A \times B$ (maximum/or), $A + B$ (minimum/and) and $\Gamma A$ (negation/not), Eqn. \ref{product}. 
\begin{equation}
\label{product}
\begin{array}{lllllll}
A \times B= \sum _i \sum_j min\{\mu_A(u_i),\mu_B(\nu_j)\}\\
A + B= \sum _i \sum_j max\{\mu_A(u_i),\mu_B(\nu_j)\}\\
\Gamma A =1-\sum _i {\mu_A(u_i)}\\
\Gamma B =1-\sum _j {\mu_B(\nu_j)}\\
~~~~~~~~~~~~~~~~~~~~~~~~~\small{i=1,2,3,...,m}\\
~~~~~~~~~~~~~~~~~~~~~~~~~\small{j=1,2,3,...,n}
\end{array}
\end{equation}
Where $m$ and $n$ are the numbers of elements in the universes of $A$ and $B$, respectively.\\
The rules are derived from the experimental findings, physiotherapy consultancy and collaboration, leading joint hypothesis (interpretation of control of human movements) and knowledge implying relations between the angles of joints \cite{dounskaia2010control}. The output is calculated based on input variables as illustrated in Fig. \ref{surface}, and defined by Eqn. \ref{Rules}. 
\begin{equation}
\label{Rules}
\begin{array}{llllll}
IF~P.E_W~is~...~AND/OR~P.E_E~is...~THEN~P.E_S~is~...\\
IF~O.E_W~is...~AND/OR~O.E_E~is...~THEN~O.E_S~is~...\\
IF~AV.E_W~is...~AND/OR~AV.E_E~is...~THEN\\~~~~AV.E_S~is~...\\ 
IF~T.E_H~is...~AND/OR~T.E_S~is...~THEN~...\\
IF~T.E_C~is...~AND/OR~T.E_R~is...~THEN~...
\end{array}
\end{equation}
\begin{figure}[!t]
\centering
\includegraphics[width=2.9in]{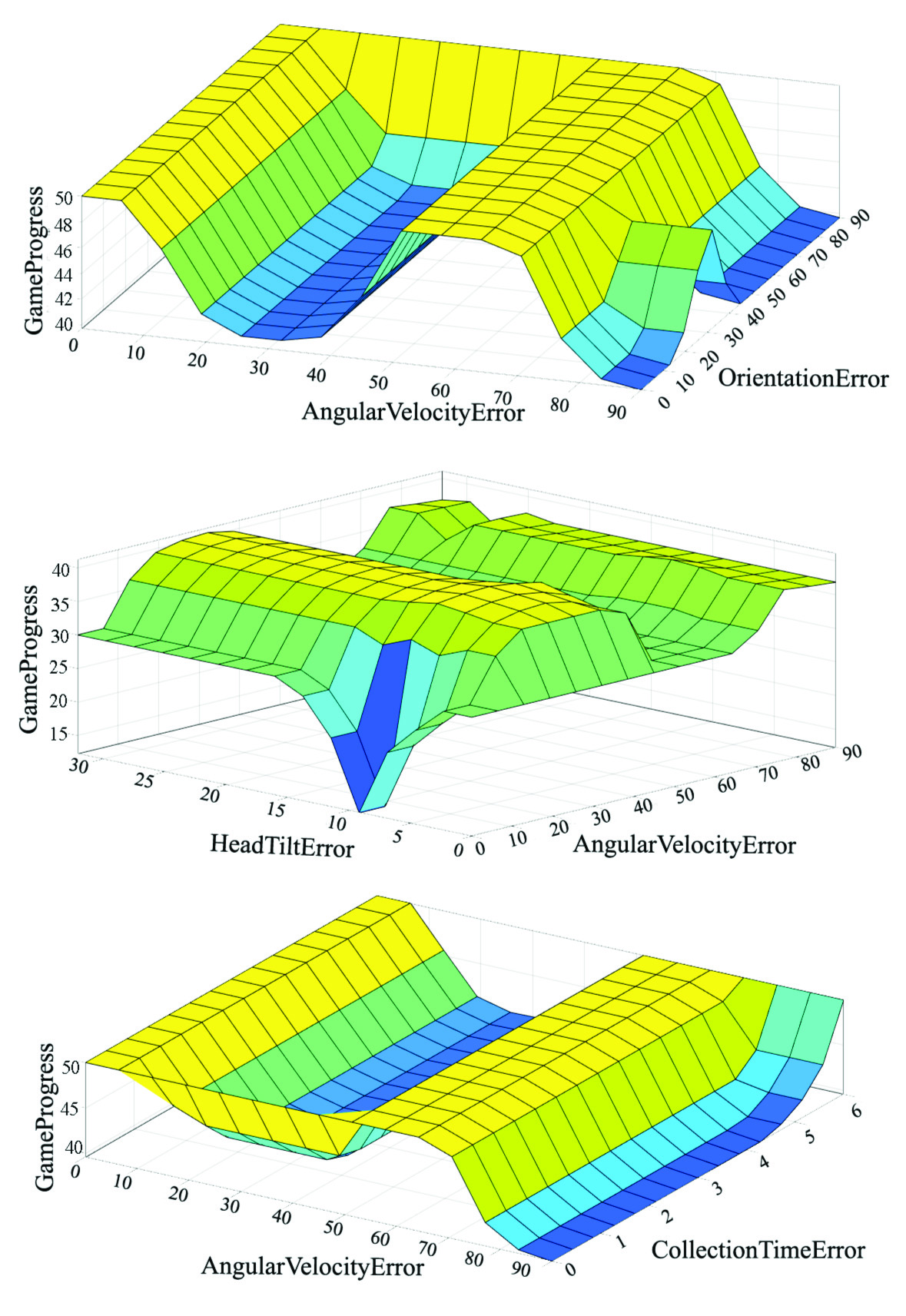}
\caption{Surface Viewer of the fuzzy inference system based on the entire span of the inputs and output in the ReHabGame.}
\label{surface}
\end{figure}

\section{Results and Discussion }
The system calculates the orientation of joints (shoulder, elbow, wrist and hand) using inverse kinematics. The posture of the body is also controlled during the game through the position and orientation of the spine to Shoulder, neck to head using fuzzy logic and rules.
Mamdani's fuzzy inference method is coded in Unity3D through a fuzzy-plugin, and the fuzzy inference process is applied at the level of individual rules. The contribution of each fuzzy set inferred is considered individually, and the final crisp control action is obtained by taking a calculus (an average) over a concrete fresh characteristic value from each one of inputs. If the input from every kinematic rule is "Very Good", with at least one input belongs to the “Good” category, then the execution of the current iteration is "Very Good". If the input from every rule is "Good", "Very Good", with at least one input belongs to the “Good” category, then the execution of that iteration is "Good". 
The developmental results show that the system allows limb movement analyses in the joint space which can be used to define define performance indicators that quantify patient residual motor capabilities and estimate the results of the therapy. 
The ReHabGame uses experience gained from exploration of a data set to improve the performance or predictive ability and provides superior predictions to guide the development of improved clinical protocols.
\section{Conclusion}
The increasing availability of 'off-the-shelf' gaming peripherals in the consumer market has provided greater opportunities for research into non-gaming or serious gaming applications. The ReHabGame developed in the current work is one example of this fusion of different technologies via relatively cheap acquisition of the Kinect Xbox, Mya armband, and Rudder Pedal devices. A fuzzy logic solution incorporated in a virtual environment is facilitated by the Unity 3D game engine. The proposed system aims to address kinematic activity of the upper and lower limbs which at the same time provides requirements to cater tasks to the needs of the patient. Various constraints, inputs and rules are employed by the player or the therapist through the menu of ReHabgame and depend on the level of the desired outcome and motor skills that need to be attained. The simulation of the upper limb motions through the robotic avatar is performed according to the position of the terminal trajectory using the IK algorithm and dynamics method. The intelligent system can halt game progression and game difficulty until the player achieves a desired outcome.
The system offers the possibility to provide a personalised, autonomously-learnt rehabilitation programme for patients with neuromuscular disorders by performing various facilitation movements.
\ifCLASSOPTIONcaptionsoff
  \newpage
\fi
\bibliographystyle{ieeetr}
\bibliography{Ref}

\begin{thebibliography}{10}

\bibitem{lai2009computational}
D.~T. Lai, R.~K. Begg, and M.~Palaniswami, ``Computational intelligence in gait
  research: a perspective on current applications and future challenges,'' {\em
  IEEE Transactions on Information Technology in Biomedicine}, vol.~13, no.~5,
  pp.~687--702, 2009.

\bibitem{xie2011iterative}
S.~Q. Xie and P.~K. Jamwal, ``An iterative fuzzy controller for pneumatic
  muscle driven rehabilitation robot,'' {\em Expert Systems with Applications},
  vol.~38, no.~7, pp.~8128--8137, 2011.

\bibitem{kavalier2005video}
F.~Kavalier, ``Video games and health,'' {\em Bmj}, vol.~331, pp.~122--3, 2005.

\bibitem{lilly2011fuzzy}
J.~H. Lilly, {\em Fuzzy control and identification}.
\newblock John Wiley \& Sons, 2011.

\bibitem{czekalski2006evolution}
P.~Czekalski, ``Evolution-fuzzy rule based system with parameterized
  consequences,'' {\em International Journal of Applied Mathematics and
  Computer Science}, vol.~16, no.~3, p.~373, 2006.

\bibitem{zarandi2008reinforcement}
M.~H.~F. Zarandi, J.~Jouzdani, and M.~F. Zarandi, ``Reinforcement learning for
  fuzzy control with linguistic states,'' {\em Journal of Uncertain Systems},
  vol.~2, no.~1, pp.~54--66, 2008.

\bibitem{zadeh1994role}
L.~A. Zadeh, ``The role of fuzzy logic in modeling, identification and
  control,'' 1994.

\bibitem{gaines1972learning}
B.~R. Gaines, ``The learning of perceptual-motor skills by men and machines and
  its relationship to training,'' {\em Instructional Science}, vol.~1, no.~3,
  pp.~263--312, 1972.

\bibitem{macdonald2007formation}
C.~MacDonald, Z.~Moussavi, and T.~Sarkodie-Gyan, ``Formation of an internal
  model of environment dynamics during upper limb reaching movements: A fuzzy
  approach,'' in {\em 2007 29th Annual International Conference of the IEEE
  Engineering in Medicine and Biology Society}, pp.~4862--4865, IEEE, 2007.

\bibitem{singh2013real}
H.~Singh, M.~M. Gupta, T.~Meitzler, Z.-G. Hou, K.~K. Garg, A.~M. Solo, and
  L.~A. Zadeh, ``Real-life applications of fuzzy logic.,'' {\em Adv. Fuzzy
  Systems}, vol.~2013, pp.~581879--1, 2013.

\bibitem{nawrocka2014fuzzy}
A.~Nawrocka, M.~Nawrocki, and A.~Kot, ``Fuzzy logic controller for
  rehabilitation robot manipulator,'' in {\em Control Conference (ICCC), 2014
  15th International Carpathian}, pp.~379--382, IEEE, 2014.

\bibitem{skoda2015estimation}
D.~Skoda, P.~Kutilek, V.~Socha, J.~Schlenker, A.~Stefek, and J.~Kalina, ``The
  estimation of the joint angles of upper limb during walking using fuzzy logic
  system and relation maps,'' in {\em Applied Machine Intelligence and
  Informatics (SAMI), 2015 IEEE 13th International Symposium on}, pp.~267--272,
  IEEE, 2015.

\bibitem{cho2015control}
J.~H. Cho, H.~S. Woo, H.~J. Lee, and C.~S. Kim, ``Control configuration for
  upper limb rehabilitation robotic systems,'' in {\em Control, Automation and
  Systems (ICCAS), 2015 15th International Conference on}, pp.~1500--1502,
  IEEE, 2015.

\bibitem{huq2013development}
R.~Huq, R.~Wang, E.~Lu, D.~H{\'e}bert, H.~Lacheray, and A.~Mihailidis,
  ``Development of a fuzzy logic based intelligent system for autonomous
  guidance of post-stroke rehabilitation exercise,'' in {\em Rehabilitation
  Robotics (ICORR), 2013 IEEE International Conference on}, pp.~1--8, IEEE,
  2013.

\bibitem{chang2008experimental}
M.-K. Chang and T.-H. Yuan, ``Experimental implementations of adaptive
  self-organizing fuzzy slide mode control to a 3-dof rehabilitation robot,''
  in {\em Innovative Computing Information and Control, 2008. ICICIC'08. 3rd
  International Conference on}, pp.~503--503, IEEE, 2008.

\bibitem{karime2014fuzzy}
A.~Karime, M.~Eid, J.~M. Alja'am, A.~El~Saddik, and W.~Gueaieb, ``A fuzzy-based
  adaptive rehabilitation framework for home-based wrist training,'' {\em IEEE
  Transactions on Instrumentation and Measurement}, vol.~63, no.~1,
  pp.~135--144, 2014.

\bibitem{fernandez2015serious}
V.~Fernandez-Cervantes, E.~Stroulia, L.~E. Oliva, F.~Gonzalez, and C.~Castillo,
  ``Serious games: Rehabilitation fuzzy grammar for exercise and therapy
  compliance,'' in {\em Games Entertainment Media Conference (GEM), 2015 IEEE},
  pp.~1--8, IEEE, 2015.

\bibitem{pirovano2012self}
M.~Pirovano, R.~Mainetti, G.~Baud-Bovy, P.~L. Lanzi, and N.~A. Borghese,
  ``Self-adaptive games for rehabilitation at home,'' in {\em 2012 IEEE
  Conference on Computational Intelligence and Games (CIG)}, pp.~179--186,
  IEEE, 2012.

\bibitem{liu2014fuzzy}
S.~Liu, Z.~Shen, M.~J. McKeown, C.~Leung, and C.~Miao, ``A fuzzy logic based
  parkinson's disease risk predictor,'' in {\em 2014 IEEE International
  Conference on Fuzzy Systems (FUZZ-IEEE)}, pp.~1624--1631, IEEE, 2014.

\bibitem{grecu2009analysis}
V.~Grecu, N.~Dumitru, and L.~Grecu, ``Analysis of human arm joints and
  extension of the study to robot manipulator,'' in {\em Proceedings of the
  International MultiConference of Engineers and Computer Scientists}, vol.~2,
  pp.~18--20, 2009.

\bibitem{papaleo2012inverse}
E.~Papaleo, L.~Zollo, S.~Sterzi, and E.~Guglielmelli, ``An inverse kinematics
  algorithm for upper-limb joint reconstruction during robot-aided motor
  therapy,'' in {\em 2012 4th IEEE RAS \& EMBS International Conference on
  Biomedical Robotics and Biomechatronics (BioRob)}, pp.~1983--1988, IEEE,
  2012.

\bibitem{tolani1996real}
D.~Tolani and N.~I. Badler, ``Real-time inverse kinematics of the human arm,''
  {\em Presence: Teleoperators \& Virtual Environments}, vol.~5, no.~4,
  pp.~393--401, 1996.

\bibitem{lura2012creation}
D.~J. Lura, {\em The Creation of a Robotics Based Human Upper Body Model for
  Predictive Simulation of Prostheses Performance}.
\newblock PhD thesis, University of South Florida, 2012.

\bibitem{welman1993inverse}
C.~Welman, {\em Inverse kinematics and geometric constraints for articulated
  figure manipulation}.
\newblock PhD thesis, Simon Fraser University, 1993.

\bibitem{esfahlani2016intelligent}
S.~S. Esfahlani and T.~Thompson, ``Intelligent physiotherapy through procedural
  content generation,'' in {\em Twelfth Artificial Intelligence and Interactive
  Digital Entertainment Conference}, 2016.

\bibitem{wang1997course}
L.-X. Wang, ``A course in fuzzy systems and control prentice hall,'' {\em
  Facsimile edition}, 1997.

\bibitem{chang2010adaptive}
M.-K. Chang, ``An adaptive self-organizing fuzzy sliding mode controller for a
  2-dof rehabilitation robot actuated by pneumatic muscle actuators,'' {\em
  Control Engineering Practice}, vol.~18, no.~1, pp.~13--22, 2010.

\bibitem{barghout2009haptic}
A.~Barghout, A.~Alamri, M.~Eid, and A.~El~Saddik, ``Haptic rehabilitation
  exercises performance evaluation using automated inference systems,'' {\em
  International Journal of Advanced Media and Communication}, vol.~3, no.~1-2,
  pp.~197--214, 2009.

\bibitem{czogala2012fuzzy}
E.~Czogala and J.~Leski, {\em Fuzzy and neuro-fuzzy intelligent systems},
  vol.~47.
\newblock Physica, 2012.

\bibitem{cordon1997applicability}
O.~Cord{\'o}n, F.~Herrera, and A.~Peregr{\'\i}n, ``Applicability of the fuzzy
  operators in the design of fuzzy logic controllers,'' {\em Fuzzy sets and
  systems}, vol.~86, no.~1, pp.~15--41, 1997.

\bibitem{alcala1999techniques}
R.~Alcal{\'a}, J.~Casillas, O.~Cord{\'o}n, F.~Herrera, and I.~Zwir,
  ``Techniques for learning and tuning fuzzy rule-based systems for linguistic
  modeling and their application,'' {\em Knowledge Engineering Systems,
  Techniques and Applications}, vol.~3, pp.~889--941, 1999.

\bibitem{bigler2012mild}
E.~D. Bigler, ``Mild traumatic brain injury: the elusive timing of
  “recovery”,'' {\em Neuroscience letters}, vol.~509, no.~1, pp.~1--4,
  2012.

\bibitem{dounskaia2010control}
N.~Dounskaia, ``Control of human limb movements: the leading joint hypothesis
  and its practical applications,'' {\em Exercise and sport sciences reviews},
  vol.~38, no.~4, p.~201, 2010.

\bibitem{siler2005fuzzy}
W.~Siler and J.~J. Buckley, {\em Fuzzy expert systems and fuzzy reasoning}.
\newblock John Wiley \& Sons, 2005.

\end{thebibliography}
\end{document}